\documentclass[a4paper,twoside,prd,preprintnumbers,longbibliography,twocolumn,nofootinbib]{revtex4-1}
\usepackage{amssymb,amsmath,bm,natbib}
\usepackage{cancel}
\usepackage[normalem]{ulem}
\usepackage{xcolor}
\usepackage{slashed}
\usepackage{graphics}
\usepackage{graphicx}
\usepackage[utf8]{inputenc}
\usepackage[T1]{fontenc}
\usepackage{tabularx}
\usepackage[caption=false]{subfig}
\usepackage{hyperref}
\usepackage{url}
\usepackage{dsfont}
\usepackage{float} 

\usepackage[uncertainty-mode=separate]{siunitx}
\usepackage{suffix} 
\newcommand{\V}[1]{\ensuremath{V_{#1}^{}}} 
\WithSuffix\newcommand\V*[1]{\ensuremath{V_{#1}^*}}

\newcommand{\DeltaM}{\Delta M_s}
\newcommand{\DeltaGamma}{\Delta \Gamma_s}
\newcommand{\asl}{a_\text{sl}^s}

\newcommand{\GeV}{{\rm GeV}}

\newcommand{\eq}[1]{Eq.~\eqref{#1}}
\newcommand{\fig}[1]{Fig.~\ref{#1}}

\renewcommand{\i}{\ensuremath{\mathrm{i}}}

\begin{document}
\preprint{PSI-PR-23-35,  ZU-TH 54/23}
\title{Diquark Explanation of \texorpdfstring{$b\to s\ell^+\ell^-$}{b -> sll}}

\author{Andreas Crivellin}
\email{andreas.crivellin@psi.ch}
\affiliation{Physik-Institut, Universit\"at Z\"urich, Winterthurerstrasse 190, CH--8057 Z\"urich, Switzerland}
\affiliation{Paul Scherrer Institut, CH--5232 Villigen PSI, Switzerland}

\author{Matthew Kirk}
\email{mjkirk@icc.ub.edu}
\affiliation{Departament de F\'isica Quàntica i Astrof\'isica (FQA), Institut de Ciències del Cosmos (ICCUB), Universitat de Barcelona (UB), Spain}

\begin{abstract}
The discrepancies between $b\to s\ell^+\ell^-$ data and the corresponding Standard Model predictions point to the existence of new physics with a significance at the $5\sigma$ level.
While previously a lepton flavour universality violating effect was preferred, the new $R(K^{(*)})$ and $B_s\to\mu^+\mu^-$ measurements are now compatible with the Standard Model, favouring a lepton flavour universal beyond the Standard Model contribution to $C_9$. Since heavy new physics is generally chiral, and because of the stringent constraints from charged lepton flavour violation, this poses a challenge for model building. In this article, we point out a novel possibility: a diquark, i.e.~a coloured scalar, induces the Wilson coefficient of the $(\bar s \gamma^\mu P_L b) (\bar c \gamma_\mu P_L c)$ operator at tree-level, which then mixes into $O_9$ via an off-shell photon penguin. This setup allows for a lepton flavour universal effect of $C_9\approx-0.5$, without violating bounds from $\DeltaM$, $\Delta\Gamma$, $B\to X_s\gamma$ and $D^0-\bar D^0$ mixing. This scenario predicts a small and negative $C_9^{\prime}$ and a light diquark, preferably with a mass around $500\,$GeV, as compatible with the CMS di-di-jet analysis, and a deficit in the inclusive $b\to c\bar c s$ rate.
\end{abstract}
\maketitle

\newpage
\section{Introduction}
While the Standard Model (SM) Cabibbo-Kobayashi-Maskawa (CKM) mechanism~\cite{Kobayashi:1973fv} of quark flavour violation was established by the $B$ factories Belle~\cite{Belle:2000cnh} and BaBar~\cite{BaBar:2001yhh}, there is still room for new physics (NP) at the order of $10\%$ in flavour changing neutral current (FCNC) processes. Such FCNC observables are loop suppressed and thus particularly sensitive to beyond-the-SM contributions. In fact, there are long-lasting hints for NP in $b\to s\ell^+\ell^-$ observables. However, the picture changed radically with the release of the latest LHCb results for the ratios $R(K^{(*)})={\rm Br}[B\to K^{(*)}\mu^+\mu^-]/{\rm Br}[B\to K^{(*)}e^+e^-]$~\cite{LHCb:2022zom,LHCb:2022qnv}, superseding their previous measurements~\cite{LHCb:2014vgu,LHCb:2017avl,LHCb:2021trn}. While previously all global fits~\cite{Altmannshofer:2021qrr,Geng:2021nhg,Alguero:2021anc,Hurth:2021nsi,Kowalska:2019ley,Ciuchini:2021smi,DAmico:2017mtc} preferred lepton flavour universality (LFU) violating NP~\cite{Crivellin:2021sff,Crivellin:2022qcj}, now data is not only consistent with LFU but even stringently limits deviations from it.

Nonetheless, the case for physics beyond the SM in $b\to s\ell^+\ell^-$ transitions remains very strong (see Ref.~\cite{Capdevila:2023yhq} for a recent review). The main tensions with the SM predictions are within the angular $B\to K^*\mu^+\mu^-$ observable $P_5^\prime$~\cite{Descotes-Genon:2012isb,LHCb:2015svh,LHCb:2020lmf,LHCb:2020gog}, the total branching ratio and angular observables in $B_s\to\phi\mu^+\mu^-$~\cite{LHCb:2015wdu,LHCb:2021zwz,LHCb:2021xxq} as well as in Br$[B\to K\mu^+\mu^-]$~\cite{LHCb:2014cxe,LHCb:2016ykl}\footnote{Measurements of these decays were also performed by the ATLAS, CMS and Belle collaborations~\cite{CMS:2013mkz,CMS:2015bcy,CMS:2017rzx,ATLAS:2018gqc,CMS:2020oqb} but with less precise results.}, with tensions at the $2-4\sigma$ level in each of these modes. In fact, while SM predictions are challenging, due to the hadronic form factors involved~\cite{Ball:2004rg,Horgan:2013hoa,Bouchard:2013eph,Bharucha:2015bzk,Gubernari:2018wyi,Cui:2022zwm} (including non-local charm-loop contributions~\cite{Khodjamirian:2010vf,Khodjamirian:2012rm,Gubernari:2020eft}), the first lattice calculation over the full $q^2$ range of Br$[B\to K\mu^+\mu^-]$ leads to a stronger tension of $4.7\,\sigma$~\cite{Parrott:2022zte}. Furthermore, $P_5^\prime$, being an optimised angular observable~\cite{Matias:2012xw,Descotes-Genon:2013vna,Descotes-Genon:2013wba} possess a reduced sensitivity to the form factors and semi-inclusive decays at high $q^2$, that are independent of hadronic form factors, are fully compatible with the other observables~\cite{Isidori:2023unk}. Finally, dispersive methods based on analyticity confirm previous error estimates for the form factors~\cite{Bobeth:2017vxj,Gubernari:2023puw} (including their non-local parts).

Combining the processes discussed above in a global fit together with all other available data on $b\to s\ell^+\ell^+$ transitions leads to a coherent picture. In fact, while before the $R(K^{(*)})$ update, the most strongly favoured scenarios were at least two-dimensional~\cite{Alguero:2019ptt}, now a single one-dimensional scenario is clearly favoured: the $C_9^U$ scenario with a significance around $5\sigma$~\cite{Gubernari:2022hxn,Ciuchini:2022wbq,Alguero:2023jeh,Wen:2023pfq}. This means that a left-handed $b-s$ current and a vectorial flavour-universal lepton current ($B_s\to\mu^+\mu^-$~\cite{LHCb:2021vsc,LHCb:2021awg,ATLAS:2018cur,CMS:2022mgd} constrains an axial current) is needed.

This poses a challenge for model building since both tree-level leptoquark effects~\cite{Hiller:2014yaa,Alonso:2015sja} as well as loop contributions of new scalars and fermions~\cite{Gripaios:2015gra,Arnan:2016cpy,Grinstein:2018fgb,Arnan:2019uhr}, in general, give a chiral lepton current and have difficulties respecting the stringent bounds from $\mu\to e$ flavour violating~\cite{Crivellin:2017dsk} unless multiple generations are involved~\cite{Crivellin:2022mff}. This leaves $Z^\prime$ models~\cite{Buras:2013qja,Gauld:2013qba,Alguero:2022wkd} as well as leptoquarks which generate a LFU effect in $C_9^U$ via a tau-loop with an off-shell photon~\cite{Crivellin:2018yvo,Crivellin:2019dwb,Fuentes-Martin:2019ign,Alguero:2022wkd,Bordone:2023ybl,FernandezNavarro:2022gst} as the remaining (simple) options. However, also in these cases $B_s-\bar B_s$ mixing~\cite{DiLuzio:2017fdq}, LEP and LHC constraints~\cite{Greljo:2022jac,Aebischer:2022oqe} make a full explanation challenging.

An alternative scenario that can naturally generate $C_9^U$ is a NP contribution to the Wilson coefficient of the $(\bar s \gamma^\mu P_L b) (\bar c \gamma_\mu P_{L,R}c)$ operator~\cite{Jager:2017gal} that mix into $C_9$ via an off-shell photon penguin~\cite{Bobeth:2014rda}. As a tree-level effect in $\bar sb\bar cc$ operators is necessary, only $Z^\prime$ bosons, heavy gluons, Higgses~\cite{Iguro:2018qzf,Crivellin:2019dun,Kumar:2022rcf,Iguro:2023jju} or diquarks (DQs) come into mind~\cite{deBlas:2017xtg}. For the first two options the possible effect is stringently limited by $B_s-\bar B_s$ mixing (to which these particles also contribute at tree-level~\cite{Aebischer:2020dsw}) and there is only a small region in parameter space left that works for 2HDMs~\cite{Iguro:2023jju}. Therefore, we will consider the DQs in this article which have not been studied so far. Out of the 8 different scalar DQs~\cite{Giudice:2011ak,Pascual-Dias:2020hxo}, there is only a single representation that couples to left-handed down-type quarks, can have flavour-diagonal couplings to up-type quarks and does not lead to tree-level effects in $\Delta F=2$: the scalar $SU(3)_c$ triplet, $SU(2)_L$ singlet with hypercharge $-1/3$ which we call $\phi$.\footnote{This field is also known as the $S_1$ leptoquark when it couples to quarks and leptons instead of two quarks. However, to avoid proton decay, we will assume that it couples only to quarks which can be achieved by e.g.~assuming that lepton number is conserved.} 

DQs are not only theoretically well motivated, e.g.~by $E_6$ models~\cite{Hewett:1988xc} or the R-parity violating MSSM~\cite{Barbier:2004ez}, but also lead to interesting LHC signatures~\cite{Ma:1998pi,Kilic:2008pm,Cheung:2002uz,Arnold:2009ay,Shu:2009xf,Dorsner:2009mq,Dorsner:2010cu,Gresham:2011dg,Patel:2011eh,Grinstein:2011yv,Ligeti:2011vt}: They are candidates from an explanation~\cite{Crivellin:2022nms} of the ATLAS di-jet~\cite{ATLAS:2021upq} and the CMS di-di-jet~\cite{CMS:2022usq} excesses. In fact, the non-resonant CMS analysis shows a weaker-than-expected limit for a diquark $SU(3)_c$ triplet with a mass around $500\,$GeV.

In the next section, we will define our model and perform the matching on the effective theory before we continue with the phenomenological analysis in Sec.~\ref{sec:pheno} can conclude in Sec.~\ref{sec:conclusions}.

\section{Setup and Observables}
\label{sec:observables}

There are 8 representations of scalar DQs with couplings to quarks $SU(2)_L$ singlets or doublets which can have either symmetric or anti-symmetric couplings in flavour space. In order to get a sizable effect in $b\to s\ell^+\ell^-$ via the operator $(\bar s \gamma^\mu P_L b) (\bar c \gamma_\mu P_{L,R}c)$, we need 1) simultaneous couplings to up-type and down-type quarks 2) flavour diagonal couplings to up-quarks (i.e.~not anti-symmetric ones) 3) left-handed down-quarks must be involved 4) no tree-level effect in $B_s-\bar B_s$ mixing. These requirements only leave a single representation, i.e.~IV in the conventions of Ref.~\cite{Giudice:2011ak} whose mass we label $M_\phi$.

The couplings to quarks are given by
\begin{equation}
    \mathcal{L} = \left(\dfrac{1}{2}\tilde \lambda_{ij}^L (\bar Q_i^{I,\alpha})^c Q_j^{J,\beta}\epsilon_{IJ} + \tilde \lambda_{ij}^R\bar u^{\alpha c}_i d^\beta_j \right)\phi^\gamma \epsilon_{\alpha \beta\gamma} + \text{h.c.} \,.
\end{equation}
Here, $\alpha,\beta,\gamma$ are color indices, $I,J$ ($i,j$) $SU(2)_L$ (flavour) indices and $c$ stands for charge conjugation. Note that $\tilde \lambda^L$ can be chosen to be symmetric in flavour space, i.e.~$\tilde \lambda_{ij}^L = \tilde \lambda^L_{ji}$, without loss of generality. After electroweak symmetry breaking, the quark doublets decompose into their $SU(2)_L$ components and in the mass eigenbasis we have
\begin{equation}
    \mathcal{L} = {\varepsilon _{\alpha \beta \gamma }}\left( {{\lambda_{ij}^L}\bar u_i^{\alpha c}{P_L}d_j^\beta  + {\lambda_{ij}^R}{{\bar u}_i^{\alpha c}}{P_R}{d_j^\beta}} \right){\phi ^\gamma }+\text{h.c.}\,,
\end{equation}
where we absorbed the rotation matrices into the definition of $\lambda^R$ and, working in the down basis, defined $V_{ii'}^*\tilde \lambda_{i'j}^L=\lambda^L_{ij}$.

\subsection{Tree-level Matching}

The Lagrangian
\begin{equation}
\mathcal{L} = -N_{ij} \sum\limits_{X = 1}^{10} {\left( {C_X^{{i}{j}}Q_X^{{i}{j}} + C_X^{\prime{i}{j}}Q_X^{\prime{i}{j}}} \right)}
\end{equation}
with $N_{ij} = 4 G_F \V{ib} \V*{js} / {{\sqrt 2 }}$ and the operators 
\begin{equation}
\begin{aligned}
Q_1^{{i}{j}} &= \left( {\bar u_i^\alpha {\gamma _\mu }{P_L}b_{}^\beta } \right)\left( {\bar s_{}^\beta {\gamma ^\mu }{P_L}u_j^\alpha } \right),\\
Q_2^{{i}{j}} &= \left( {\bar u_i^\alpha {\gamma _\mu }{P_L}b_L^\alpha } \right)\left( {\bar s_L^\beta {\gamma ^\mu }{P_L}u_j^\beta } \right),\\
Q_7^{{i}{j}} &= \left( {\bar u_i^\alpha {P_R}b_{}^\beta } \right)\left( {\bar s_{}^\beta {P_R}u_j^\alpha } \right),\\
Q_8^{{i}{j}} &= \left( {\bar u_i^\alpha {P_R}b_{}^\alpha } \right)\left( {\bar s_L^\beta {P_R}u_j^\beta } \right),\\
Q_9^{{i}{j}} &= \left( {\bar u_i^\alpha {\sigma _{\mu \nu }}{P_R}b^\beta } \right)\left( {\bar s_{}^\beta {\sigma ^{\mu \nu }}{P_R}u_j^\alpha } \right),\\
Q_{10}^{{i}{j}} &= \left( {\bar u_i^\alpha {\sigma _{\mu \nu }}{P_R}b^\alpha } \right)\left( {\bar s_L^\beta {\sigma ^{\mu \nu }}{P_R}u_j^\beta } \right),
\end{aligned}
\end{equation}
defines the charged current interactions. Our diquark, integrated out at tree level at the electroweak scale, leads to the coefficients
\begin{align}
C_1^{{i}{j}} &=  - C_2^{{i}{j}} = -\dfrac{{\lambda _{i2}^{L*}\lambda _{j3}^L}}{{2N_{ij}M_\phi ^2}},\\
C_7^{{i}{j}} &=  - C_8^{{i}{j}} =  - 4C_9^{{i}{j}} = 4C_{10}^{{i}{j}} = \dfrac{{\lambda _{i2}^{L*}\lambda _{j3}^R}}{{2N_{ij}M_\phi ^2}},
\end{align}
with the primed coefficients obtained by ${\lambda^L \leftrightarrow \lambda^R}$.\footnote{It can be seen that $C_X^{uc}$ (which give $b \to u \bar{c} s$ transitions) are enhanced relative to the SM by the large ratio $\V{us} / \V{ub}$. However, the effects of these operators in meson mixing-related observables is too small~\cite{Lenz:2022pgw}.
Furthermore, $C^{cu}_X$ (which give $b \to c \bar{u} s$ transitions) have been considered as part of a potential explanation of the discrepancy between the SM prediction with QCD factorization and experiment for $\bar{B}^0_{(s)} \to D^{*+}_{(s)} \{\pi^-, K^-\}$~\cite{Huber:2016xod,Bordone:2020gao,Cai:2021mlt,Endo:2021ifc,Beneke:2021jhp,Fleischer:2021cct,Fleischer:2021cwb,Piscopo:2023opf}. However as no analysis has been done with just NP in $\bar s b \bar c u$, this interesting future direction is beyond the scope of this article.}

\subsection{\texorpdfstring{$b\to s\ell^+\ell^-$}{b -> sll}}

The effective Lagrangian governing $b \to s \ell \ell$ transitions is given by
\begin{equation}
\mathcal{L} \supset N_{33} \frac{\alpha}{4\pi}
\begin{aligned}[t]
    &\left[ C_9 (\bar s \gamma^\mu P_L b)(\bar \ell \gamma_\mu \ell) + C_{10} (\bar s \gamma^\mu P_L b)(\bar \ell \gamma_\mu \gamma^5 \ell) \right],
\end{aligned}
\end{equation}
where the primed operators and coefficients are again obtained by exchanging $L$ and $R$. The tree-level induced operators $Q_{1,2}^{cc}$ generate via mixing
\begin{equation}
\begin{aligned}[t]
C_9 (m_b) = &8.5 C_1^{cc} (M_W) + 2.1 C_2^{cc} (M_W) 
\\
+ &\left[3.1 C_1^{cc} (M_W) + 0.32 C_2^{cc} (M_W) \right] \times h(q^2),
\end{aligned}
\end{equation}
at the $B$ meson scale (and similarly for $C_9^\prime$)
with
\begin{equation}
h(q^2, m_c) = -\frac{4}{9} \left[ \ln \frac{m_c^2}{m_b^2} - \frac{2}{3} + (2 + z) a(z) - z  \right],
\end{equation}
$a(z)= \sqrt{|z-1|} \arctan \frac{1}{\sqrt{z -1}}$, and $z = 4 m_c^2/q^2$. Note that $h$ also includes finite subleading $q^2$-dependent effects (which we evaluate at $q^2 = \qty{5}{\GeV^2}$) \cite{Jager:2017gal}.

The threshold effects from top quark loops induce at the EW scale
\begin{align}
\frac{{-C_9}}{1 - 4s_W^2}={C_{10}} &=  - \frac{{\lambda _{32}^{L*}\lambda _{33}^L}}{{2{e^2}V_{tb}^{}V_{ts}^*}}f\left( {\frac{{m_t^2}}{{{M_\phi^2}}}} \right),
\end{align}
with $f(x) = x ( x - \log x - 1) / (x - 1)^2$ and $C_{9,10}^\prime$ obtained from $C_{9,10}$ by exchanging $L$ and $R$. 

\subsection{\texorpdfstring{$B \to X_s \gamma$}{B -> Xs gamma}}

The relevant effective operators are defined by
\begin{equation}
\mathcal{L} \supset N_{33}  C_{7\gamma (8g)} \frac{e (g_s) m_b}{16 \pi^2} \bar{s} \sigma^{\mu \nu} (t^a) P_R b F_{\mu \nu} (G^a_{\mu \nu}) ,
\end{equation}
with primed operators obtained through $P_R \to P_L$. We have two NP contributions to $C_{7\gamma}$. The mixing of the $\bar s b \bar c c$ operators into $C_{7\gamma}$ leads to~\cite{Jager:2017gal,Jager:2019bgk}
\begin{align}
C_{7\gamma} (m_b) 
\begin{aligned}[t]
=&0.02 C_1^{cc} (M_W) -0.19 C_2^{cc} (M_W) -1.0 C_7^{cc} (M_W) 
\\
-&0.47 C_8^{cc} (M_W) + 4.0 C_9^{cc} (M_W) +0.47C_{10}^{cc} (M_W)
\\
-&\frac{m_c}{m_b} (2.5 C_7^{cc} +  1.3 C_8^{cc} - 10 C_9^{cc}  + 0.89 C_{10}^{cc}) \times y,
\end{aligned}
\end{align}
where, similarly to the case of $C_9$, we have included sub-leading $q^2$ dependent terms via
$y = -(1 + 2 \log (m_c^2 / m_b^2))/6$. The equivalent result for the primed operators is again obtained by an exchange of chiralities.\footnote{In principle, there are potential RGE contributions from $\bar s b \bar cc$ operators in $C_{8g}$ which are however expected to be very small (see appendix of Ref.~\cite{Jager:2017gal}).}

Second, the direct matching at the new physics scale (to be taken the weak scale) gives direct contributions to $C_{7\gamma,8g}^{(\prime)} (M_W)$.
The most important result is the existence of $m_t/m_b$ enhanced terms for both coefficients which are proportional to $\lambda^L_{22} \lambda^R_{33}$, while the full expressions can be found in the supplementary material~\cite{Supp}.
These coefficients are then evolved to the $B$ meson scale resulting in 
\begin{equation}
C_{7{\gamma}}^{(\prime)} (m_b) = 0.65 C_{7\gamma}^{(\prime)} (M_W) + 0.10 C_{8g}^{(\prime)} (M_W)\,.
\end{equation}

The latest SM prediction for the inclusive radiative decay $B \to X_s \gamma$~\cite{Misiak:2020vlo}
\begin{equation}
\text{Br}[B \to X_s \gamma]^{\rm SM} = \num{3.40 \pm 0.17 e-4} \,,
\end{equation}
is in good agreement with the experimental average~\cite{HeavyFlavorAveragingGroup:2022wzx}
\begin{equation}
\text{Br}[B \to X_s \gamma]^{\rm EXP} = \num{3.49 \pm 0.19 e-4} \,.
\end{equation}
This measurement stringently constrains a BSM contribution to $C_{7\gamma}$. In addition, asymmetries in $B \to K^* e^+e^-$~\cite{LHCb:2020dof} are very sensitive to $C_{7\gamma}^\prime$. We performed a fit to $C_{7\gamma}$ and $C_{7\gamma}^\prime$ using \texttt{smelli}~\cite{Stangl:2020lbh,Aebischer:2018iyb,Straub:2018kue}.

\subsection{\texorpdfstring{$B_s-\bar B_s$}{Bs} Mixing}

In the SM, EW box diagrams give rise to $B_s-\bar B_s$ mixing, which can be measured through observables including $\DeltaM$, $\DeltaGamma$, and $\asl$ that can in principle constrain our NP model.

The mass difference $\DeltaM = 2 |M_{12}^s|$ in the SM is
\begin{equation}
\DeltaM^\text{SM} = (18.2^{+0.6}_{-0.8}) \, \text{ps}^{-1} \,,
\end{equation} 
using the non-perturbative bag parameter combination from Ref.~\cite{Greljo:2022jac} (based on results in Refs.~\cite{FermilabLattice:2016ipl,Dowdall:2019bea,King:2019lal}), the FLAG 2023 average for $f_{B_s}$ from FLAG 2023~\cite{FlavourLatticeAveragingGroupFLAG:2021npn} (based on results in Refs.~\cite{Dowdall:2013tga,ETM:2016nbo,Hughes:2017spc}), $m_t$ from PDG 2022~\cite{ParticleDataGroup:2022pth}, and the Spring 2021 CKMfitter collaboration results~\cite{Charles:2004jd,CKMfitter:Spring2021}, which should be compared to the latest HFLAV average of~\cite{HeavyFlavorAveragingGroup:2022wzx} 
\begin{equation}
\DeltaM^{\rm EXP} = \qty{17.765 \pm 0.006}{\per \ps}\,.
\end{equation}

Our new physics contributes through various 1-loop box diagrams, which generate electroweak scale $\Delta B=2$ coefficients whose full expressions are given in the supplementary material \cite{Supp}.

The width difference for $B_s$ mesons can be calculated as $\DeltaGamma = 2 |\Gamma_{12}^s| \cos( \arg (-M_{12}^s / \Gamma_{12}^s))$, where the current SM prediction is~\cite{Lenz:PrivateFPCP2023}
\begin{equation}
\DeltaGamma = \qty{0.0895 \pm 0.0131}{\per \ps} \,,
\end{equation}
and the latest HFLAV average\footnote{There is some tension between the ATLAS, CMS, and LHCb measurements, such that this HLFAV average applied a scale factor of 1.8 (according to the PDG prescription) to the total uncertainty.} is~\cite{HeavyFlavorAveragingGroup:2022wzx}
\begin{equation}
\DeltaGamma = \qty{0.083 \pm 0.005}{\per \ps} \,.
\end{equation}
Our model alters this quantity through the $\bar s b\bar cc$ operators, where the full NP contributions were calculated in Refs.~\cite{Jager:2017gal,Jager:2019bgk}.

In addition the semi-leptonic asymmetry $\asl$ receives a large NP contribution since our NP does not suffer the severe GIM cancellation seen in the SM.
However, at least orders of magnitude improvement in the experimental precision would be required for this effect to be observable, and so we make no further mention of it here (some more detailed discussion can be found in the supplementary material \cite{Supp}).

\subsection{\texorpdfstring{$D^0-\bar D^0$}{D0} mixing}

The $\Delta C=2$ coefficients which give a short-distance contribution to $\Delta M_D$ are
\begin{align}
{C_1} &= \frac{1}{64 \pi^2 M_\phi^2}  \left[ \left( \lambda^L \lambda^{L \dagger} \right)_{12} \right]^2,
\\
{C_2} &= {C_3} = 0,
\\
{C_4} &= C_5 = -\frac{1}{32 \pi^2 M_\phi^2} \left( \lambda^L \lambda^{L \dagger} \right)_{12} \left( \lambda^R \lambda^{R \dagger} \right)_{12} \,.
\end{align}
$C_{1,2,3}^{\prime}$ are found from $C_{1,2,3}$ by exchanging $\lambda^L \leftrightarrow \lambda^R$.

The SM prediction is currently unclear, as a naive calculation gives a result four or five orders of magnitude too small, while other estimates (albeit not from first principles) suggest the SM alone could give a result $x^\text{SM} \sim \qty{0.1}{\percent}$ \cite{Wolfenstein:1985ft,Donoghue:1985hh,Falk:2001hx,Falk:2004wg,Cheng:2010rv,Jiang:2017zwr,Lenz:2020efu}, where 
\begin{equation}
x \equiv \frac{2 \Delta M_D}{\Gamma_D} ,
\end{equation}
is the commonly reported observable in $D^0-\bar D^0$-mixing. Comparing this to the current HLFAV average \cite{HeavyFlavorAveragingGroup:2022wzx}
\begin{equation}
x^\text{\rm EXP} = \qty{0.407 \pm 0.044}{\percent},
\end{equation}
we take a conservative approach and allow the short-distance NP contribution to be up to twice the size of the experimental value (i.e.\ we impose $\Delta M_D^\text{NP} \leq 2 \Delta M_D^\text{exp}$, allowing in principle for a \qty{100}{\percent} cancellation between BSM and SM).

\subsection{$B_s / B_d$ lifetime ratio}

The ratio of the $B_s$ to $B_d$ lifetimes was long thought to be a theoretically clean observable, benefiting from many cancellations of uncertainties.
However, recent calculations of the SM contribution to the Darwin operator~\cite{Lenz:2020oce,Mannel:2020fts,Moreno:2020rmk} has lead to a situation where the theory prediction is unclear, and in addition there is some tension between the different experimental measurements \cite{HeavyFlavorAveragingGroup:2022wzx} (see the supplementary material \cite{Supp} for further discussion).
As such we do not consider this observable further.

\section{Phenomenology}
\label{sec:pheno}

We now turn to the phenomenology of our model. First of all, we set $M_\phi=500\,$GeV which is compatible with the non-resonant paired di-jet search of CMS due to the weaker-than-expected limit in this mass region~\cite{CMS:2022usq}. Note that using a light DQ helps to reduce the relative effect in $\Delta F=2$ processes since here the DQ contribution is proportional to $\lambda^4/M_\phi^2$ while for all other flavour processes the leading DQ effect has a $\lambda^2/M_\phi^2$ scaling. 

The product of $\tilde\lambda^L_{23}$ and $\tilde\lambda^L_{22}$ is necessary to give the effect in $b\to s\ell^+\ell^-$ via $C_9$ while the product of $\tilde{\lambda}^R_{22}$ and  $\tilde{\lambda}^R_{23}$ ($\lambda^R_{12}$) helps to weaken the bound from $B_s-\bar B_s$ mixing ($D^0-\bar D^0$). To avoid a chirally enhanced effect in $b\to s\gamma$ $\lambda^R_{33}\approx0$ is helpful. Furthermore, to avoid bounds from di-jet resonance searches~\cite{ATLAS:2018qto,Bordone:2021cca} and Kaon mixing, we assume the left-handed coupling involving the first generation to be approximately zero. Thus we consider the following structure for the DQ quark couplings
\begin{equation}
\tilde{\lambda}^L = 
\begin{pmatrix}
0 & 0 & 0 \\
0 & \tilde{\lambda}^L_{22} & \tilde{\lambda}^L_{23} \\
0 & \tilde{\lambda}^L_{23} & 0
\end{pmatrix}
\,,\qquad \lambda^R = 
\begin{pmatrix}
0 & \lambda^R_{12} & 0 \\
0 & \lambda^R_{22} & \lambda^R_{23} \\
0 & 0 & 0
\end{pmatrix}\,.
\end{equation}
Note that we input ${\tilde\lambda}^L$ in the down quark basis (not including CKM rotations). 

\begin{figure*}[t]
    \includegraphics[width=\textwidth]{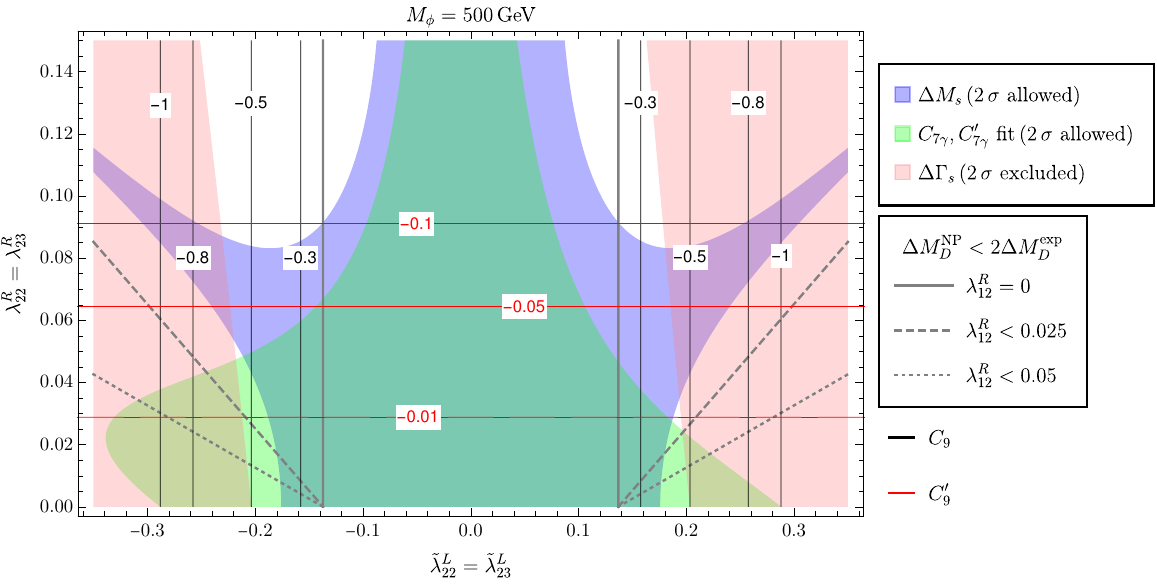}
    \caption{Experimental constraints given our assumed UV couplings matrices. The black and red vertical and horizontal lines show the generated LFU contributions to $C_9$ and $C_9^\prime$, respectively.}
    \label{fig:mainplot}
\end{figure*}

Since $\tilde\lambda^{L*}_{22}\tilde\lambda^L_{23}$ must be positive to give the preferred sign in $C_9$ we set for simplicity $\tilde{\lambda}^L_{22} = \tilde{\lambda}^L_{23}$ and $\lambda^R_{22} = \lambda^R_{23}$, assume real couplings and show the preferred regions of the various observables 
in \fig{fig:mainplot}. We see that a reasonably sizable LFU $C_9$, of the order of $-0.5$, can be generated in our model while still respecting the other experimental constraints.
For $D^0-\bar D^0$-mixing, we show the regions compatible with our fine-tuning argument for $\lambda^R_{12}=0$ and floating it within two different (small) ranges which are compatible with LHC di-jet searches. Interestingly, our model predicts $|C_9^\prime|\ll |C_9|$ but with the same sign as (slightly) preferred by the current global fit~\cite{Alguero:2023jeh}. Note that generating the sizeable negative $C_9$ leads to a positive shift in $\DeltaGamma$. This is more in line with the latest experimental result from CMS, which is in slight tension with LHCb and ATLAS measurements \cite{HeavyFlavorAveragingGroup:2022wzx}.

\section{Conclusions and Outlook}
\label{sec:conclusions}

There are persistent and significant tensions in $b\to s\ell^+\ell^-$ observables. They are most pronounced in Br[$B\to K \mu^+\mu^-$], in the angular observable $P_5^\prime$ (in $B\to K^*\mu^+\mu^-$) and the total branching ratio as well as angular observables in $B_s\to \phi\mu^+\mu^-$. In combination with the constraints from $R(K^{(*)})$ and $B_s\to \mu^+\mu^-$ they point towards lepton flavour universal NP in $C_9$ with a significance at the $5\sigma$ level.
This poses a challenge for model building because heavy NP is generally chiral, while for generating a dominant effect in $C_9$, a vectorial lepton current is needed. Furthermore, bounds from charged lepton flavour violation require a separation of the electron and muon sectors, which is difficult to achieve in many models. 

In this article, we proposed a novel model explaining the $b\to s\ell^+\ell^-$ anomalies: an $SU(3)_c$ triplet scalar DQ with hypercharge $Y=-1/3$. This field can generate a LFU effect in $C_9$ via the mixing of the $(\bar s \gamma^\mu P_L b)(\bar c \gamma_\mu P_L c)$ operator into $O_9$. Since only quarks are directly involved in this model, charged lepton flavour violation is automatically respected. Furthermore, due to the weaker-than-expected LHC limit, DQs can still be relatively light, around 500\,GeV.

We show that in this setup one can obtain $C_9\approx-0.5$ while respecting the bounds from $B_s-\bar B_s$ and $D^0-\bar D^0$ mixing as well as LHC searches. We predict a small negative value of $C_9^\prime$ as well as a positive shift in $\Delta\Gamma_s$ of around \qty{20}{\percent} w.r.t.~the SM prediction. For $b\to s\gamma$ our model predicts NP shifts in $|C_{7\gamma}^{(\prime)}|\approx0.05$.
Furthermore, DQs are candidates for sizable effects in hadronic decays of mesons, like $\epsilon^\prime/\epsilon$~\cite{Aebischer:2020jto,Buras:2020wyv} and could explain several anomalies in non-leptonic $B$ decays like the longitudinal polarization in $B_q\to K^*\bar K^*$~\cite{Alguero:2020xca}, the $B\to K\pi$ puzzle~\cite{Hofer:2010ee,Fleischer:2018bld,Crivellin:2019isj}, the $B_{(s)} \to D_{(s)} \{ K,\pi\}$~\cite{Huber:2016xod,Bordone:2020gao,Cai:2021mlt,Endo:2021ifc,Beneke:2021jhp,Fleischer:2021cct,Fleischer:2021cwb} discrepancy within QCD factorisation or the LHC di-di-jet excess~\cite{Crivellin:2022nms}.

Finally, our NP contributions to the $C_{1,2}^{cc}$ coefficients have opposite sign than $C_1^{cc}$ in the SM. This reduces the theory prediction for the inclusive branching ratio $b \to c \bar c s$ where in the SM we have Br$[b\to c\bar c s]_{\rm SM}=23\pm2$\%~\cite{Krinner:2013cja}. However, while the situation for the experimental determination of this fully inclusive quantity is currently quite unclear, and $O(30\%)$ effects in Br$[b\to c\bar c s]$ are possible, our model prediction is in line with the so-called ``missing charm puzzle''~\cite {Lenz:2000kv}. A clarification of the experimental situation would therefore be of great interest.

\begin{acknowledgements}
The work of A.C.~is supported by a Professorship Grant (PP00P2\_211002) of the Swiss National Science Foundation. M.K.\ acknowledges support from a Maria Zambrano fellowship, from the State Agency for Research of the Spanish Ministry of Science and Innovation through the ``Unit of Excellence María de Maeztu 2020-2023'' award to the Institute of Cosmos Sciences (CEX2019-000918-M), and from PID2019-105614GB-C21 and 2017-SGR-929 grants.
\end{acknowledgements}

\bibliography{bibliography}

\onecolumngrid

\setcounter{equation}{0}
\setcounter{figure}{0}
\setcounter{table}{0}
\setcounter{section}{0}
\makeatletter
\renewcommand{\theequation}{S.\arabic{section}.\arabic{equation}}
\renewcommand{\thefigure}{S.\arabic{figure}}
\renewcommand{\thetable}{S.\arabic{table}}
\renewcommand{\thesection}{\arabic{section}}

\newpage
\begin{center}
 \large{\bf Supplemental Material}
 \end{center}

\section{Wilson coefficients}

\subsection{$B_s$ mixing}
The $B_s$ mixing wilson coefficients generated at the EW scale are given by
\begin{align}
{C_1} &= \dfrac{{ - 1}}{{64{\pi ^2}}}\sum\limits_{j,k = 1}^3 {\lambda _{j2}^{L*}} \lambda _{j3}^L\lambda _{k2}^{L*}\lambda _{k3}^L{D_2}\left( {m_{{u_j}}^2,m_{{u_k}}^2,M_\phi ^2,M_\phi ^2} \right),
\\
{C_2} &= {C_3} = \dfrac{{ - 1}}{{32{\pi ^2}}}\sum\limits_{j,k = 1}^3 {{m_{{u_j}}}} {m_{{u_k}}}\lambda _{j2}^{R*}\lambda _{j3}^L\lambda _{k2}^{R*}\lambda _{k3}^L{D_0}\left( {m_{{u_j}}^2,m_{{u_k}}^2,M_\phi ^2,M_\phi ^2} \right),
\\
{C_4} &= \dfrac{{ - 1}}{{16{\pi ^2}}}\sum\limits_{j,k = 1}^3 
\begin{aligned}[t]
&\left( 
{m_{{u_j}}}{m_{{u_k}}}\lambda _{j2}^{R*}\lambda _{j3}^L\lambda _{k2}^{L*}\lambda _{k3}^R{D_0}\left( {m_{{u_j}}^2,m_{{u_k}}^2,M_\phi ^2,M_\phi ^2} \right) 
 - \frac{1}{2}\lambda _{j2}^{R*}\lambda _{j3}^R\lambda _{k2}^{L*}\lambda _{k3}^L{D_2}\left( {m_{{u_j}}^2,m_{{u_k}}^2,M_\phi ^2,M_\phi ^2} \right)
\right.
\\
&\left. \; + m_{u_j} m_{u_k} \V{j3} \V*{k2} \lambda^R_{j3} \lambda^{R*}_{k2}\left[ g^2 D_0\left(m_{u_j}^2, m_{u_k}^2, M_W^2, M_\phi^2 \right) - \sqrt{2} G_F D_2\left(m_{u_j}^2, m_{u_k}^2, M_W^2, M_\phi^2 \right) \right] \right) ,
\end{aligned}
\\
{C_5} &= \dfrac{{ - 1}}{{16{\pi ^2}}}\sum\limits_{j,k = 1}^3\begin{aligned}[t]
&\left( 
{m_{{u_j}}}{m_{{u_k}}}\lambda _{j2}^{R*}\lambda _{j3}^L\lambda _{k2}^{L*}\lambda _{k3}^R{D_0}\left( {m_{{u_j}}^2,m_{{u_k}}^2,M_\phi ^2,M_\phi ^2} \right) 
 - \frac{1}{2}\lambda _{j2}^{R*}\lambda _{j3}^R\lambda _{k2}^{L*}\lambda _{k3}^L{D_2}\left( {m_{{u_j}}^2,m_{{u_k}}^2,M_\phi ^2,M_\phi ^2} \right)
\right.
\\
&\left. \; -  m_{u_j} m_{u_k} \V{j3} \V*{k2} \lambda^R_{j3} \lambda^{R*}_{k2} \left[ g^2 D_0\left(m_{u_j}^2, m_{u_k}^2, M_W^2, M_\phi^2 \right) - \sqrt{2} G_F D_2\left(m_{u_j}^2, m_{u_k}^2, M_W^2, M_\phi^2 \right) \right] \right) .
\end{aligned}
\end{align}
$C_{1,2,3}^{\prime}$ are obtained by exchanging $\lambda^L \leftrightarrow \lambda^R$ in $C_{1,2,3}$, and the effective Hamiltonian and the loop functions are defined in Appendix A and C of Ref.~\cite{Crivellin:2013wna}.

\subsection{\texorpdfstring{$b \to s \gamma$}{b to s gamma}}
The dipole coefficients generated at the EW scale which are relevant for $b \to s \gamma$ are given by 
\begin{align}
{C_{7\gamma}} (M_W) &= \frac{{ - 1}}{{M_\phi ^2}} 
\frac{2}{N_{33}}
\left( {\lambda _{j2}^{L*}\lambda _{j3}^R\frac{{{m_{{u_j}}}}}{{{m_b}}} \left[ \frac{1}{3} {{f_\Phi }\left( {\frac{{m_{{u_j}}^2}}{{M_\phi ^2}}} \right) - \frac{2}{3}{g_\Phi }\left( {\frac{{m_{{u_j}}^2}}{{M_\phi ^2}}} \right)} \right] + \lambda _{j2}^{L*}\lambda _{j3}^L\left[ \frac{1}{3} {{{\tilde f}_\Phi }\left( {\frac{{m_{{u_j}}^2}}{{M_\phi ^2}}} \right) - \frac{2}{3}{{\tilde g}_\Phi }\left( {\frac{{m_{{u_j}}^2}}{{M_\phi ^2}}} \right)} \right]} \right),\nonumber
\\
C_{8g} (M_W) &= \frac{-1}{{M_\phi ^2}}
\frac{2}{N_{33}}
\left( {\lambda _{j2}^{L*}\lambda _{j3}^R\frac{{{m_{{u_j}}}}}{{{m_b}}} \left[ -2 f_\Phi \left( \frac{m_{u_j}^2}{M_\phi^2} \right) + {g_\Phi }\left( {\frac{{m_{{u_j}}^2}}{{M_\phi ^2}}} \right) \right] + \lambda _{j2}^{L*}\lambda _{j3}^L \left[ -2 \tilde{f}_\Phi \left( \frac{m_{u_j}^2}{M_\phi^2} \right) + {{\tilde g}_\Phi }\left( {\frac{{m_{{u_j}}^2}}{{M_\phi ^2}}} \right) \right]} \right),
\end{align}
with the loop functions defined in Ref.~\cite{Crivellin:2018qmi}, and primed operators obtained as usual via $\lambda^L \leftrightarrow \lambda^R$.

\section{$B_s / B_d$ lifetime ratio}
The lifetime ratio $\tau(B_s)/\tau(B_d)$ has long been considered a theoretically clean observable, which benefits from many cancellations of uncertainties in the ratio.
The current status is however more complicated.
On the the experimental side, the 2023 HFLAV average is~\cite{HeavyFlavorAveragingGroup:2022wzx}
\begin{equation}
\frac{\tau(B_s)^{\rm EXP}}{\tau(B_d)^{\rm EXP}} = \num{1.002 \pm 0.004} \,,
\end{equation}
yet in computing this average the combination has had scale factor of 2.6 applied to the total error to account for a tension between the ATLAS, LHCb and CMS measurements of the $B_s$ lifetime.
One recent SM prediction is~\cite{Lenz:2022rbq} 
\begin{equation}
\label{eq:lifetime_ratio_SM_A}
\frac{\tau(B_s)^{\rm SM}}{\tau(B_d)^{\rm SM}} = \num{1.003 \pm 0.006} \,,
\end{equation}
however, the uncertainty of the experimental extraction of the non-hadronic Darwin term matrix element leads to an alternative SM prediction of
\begin{equation}
\label{eq:lifetime_ratio_SM_B}
\frac{\tau(B_s)^{\rm SM}}{\tau(B_d)^{\rm SM}} = \num{1.028 \pm 0.011} \,,
\end{equation}
which shows a much larger tension with experiment (see Ref.~\cite{Lenz:2022rbq} for more discussion).
In this second case, in our model there are potentially sizable, but currently unknown, contributions from $C_{7-10}^{cc}$ to the Darwin operator coefficient which renders the theoretical prediction incomplete (the contribution from the SM operators $C_{1,2}^{cc}$ has only recently been calculated in Refs.~\cite{Lenz:2020oce,Mannel:2020fts,Moreno:2020rmk}).

When relying on the SM prediction in \eq{eq:lifetime_ratio_SM_A}, we find that the lifetime ratio observable does not provide a useful bound on our model.
On the other hand, the tension with experiment of the alternative SM prediction given in \eq{eq:lifetime_ratio_SM_B} cannot be resolved in our model without inducing sizable discrepancies elsewhere.

\section{$a_\text{sl}^s$}

The semi-leptonic asymmetry for $B_s$ mesons can be calculated as $\asl = \Im (\Gamma_{12}^s / M_{12}^s)$, where the current SM prediction is~\cite{Lenz:PrivateFPCP2023}
\begin{equation}
a_\text{sl}^{s, \text{SM}} = \num{2.2 \pm 0.3 e-5} \,,
\end{equation}
which should be compared to the latest HFLAV average of~\cite{HeavyFlavorAveragingGroup:2022wzx} 
\begin{equation}
a_\text{sl}^{s, \text{EXP}} = \num{-60 \pm 280 e-5} \,.
\end{equation}
As our NP does not have the severe GIM cancellation of the SM contribution to $\Gamma_{12}^s$ (see the discussion in e.g.\ Ref.~\cite{Artuso:2015swg}), we will predict an asymmetry that is twice the size, but with opposite sign, as the expected SM value (i.e. $a_\text{sl}^{s, \text{NP}} \approx \num{-4e-5}$), even without CP violating couplings, for the parameter space of our model which gives $C_9$ of the order of $-0.5$.
However, as can be seen by comparing the experimental precision with the theoretical predictions, differentiating our NP value from the SM result would require a improvement in the experimental precision of around two orders of magnitude, and so this observable is not currently a useful indicator in our case.

\end{document}